\documentclass[11pt]{article}

\usepackage{amsthm}
\usepackage{amsmath}
\usepackage{amssymb}
\usepackage{bm}
\usepackage{mathrsfs}
\usepackage[dvips]{graphicx}
\usepackage{algorithmic}
\usepackage{algorithm}
\usepackage{color}
\usepackage{url}
\usepackage{supertabular}
\usepackage[tmargin=1in,bmargin=1in,lmargin=1in,rmargin=1in]{geometry}
\usepackage{fancyhdr}
\usepackage[english]{babel}
\usepackage[utf8]{inputenc}
\usepackage{amsmath}
\usepackage{graphicx}
\usepackage{natbib}

\newcommand{\be}{\begin{equation}}
\newcommand{\ee}{\end{equation}}
\newcommand{\bes}{\begin{equation*}}
\newcommand{\ees}{\end{equation*}}
\newcommand{\beqn}{\begin{eqnarray}}
\newcommand{\eeqn}{\end{eqnarray}}
\newcommand{\beqns}{\begin{eqnarray*}}
\newcommand{\eeqns}{\end{eqnarray*}}
\newcommand{\lkr}{\left(}
\newcommand{\lkv}{\left[}
\newcommand{\rkv}{\right]}
\newcommand{\rkr}{\right)}
\newcommand{\lfi}{\left\{}
\newcommand{\rfi}{\right\}}

\newcommand{\fr}[1]{(\ref{#1})}

\newcommand{\del}{\delta}
\newcommand{\Del}{\Delta}

\newcommand{\al}{\alpha}

\newcommand{\eps}{\epsilon}

\newcommand{\om}{\omega}
\newcommand{\lam}{\lambda}

\newcommand{\Om}{\Omega}
\newcommand{\Sig}{\Sigma}

\newcommand{\II}{\ensuremath{{\mathbb I}}}

\newcommand{\RR}{{\mathbb R}}

\newcommand{\etal}{{\it et  al. }}

\newtheorem{theorem}{Theorem}
\newtheorem{lemma}{Lemma}
\newtheorem{corollary}{Corollary}
\newtheorem{proposition}{Proposition}
\newtheorem{remark}{Remark}

\newcommand{\boe}{\mathbf{e}}

\newcommand{\bt}{\mathbf{t}}

\newcommand{\bx}{\mathbf{x}}
\newcommand{\by}{\mathbf{y}}

\newcommand{\bA}{\mathbf{A}}
\newcommand{\bB}{\mathbf{B}}

\newcommand{\bD}{\mathbf{D}}

\newcommand{\bJ}{\mathbf{J}}

\newcommand{\bO}{\mathbf{O}}
\newcommand{\bP}{\mathbf{P}}
\newcommand{\bQ}{\mathbf{Q}}
\newcommand{\bH}{\mathbf{H}}
\newcommand{\bU}{\mathbf{U}}

\newcommand{\bX}{\mathbf{X}}
\newcommand{\bY}{\mathbf{Y}}

\newcommand{\bzero}{\mathbf{0}}
\newcommand{\bone}{\mathbf{1}}

\newcommand{\bLam}{\mbox{\mathversion{bold}$\Lambda$}}

\newcommand{\bTe}{\mbox{\mathversion{bold}$\Theta$}}

\newcommand{\calD}{{\mathcal{D}}}
\newcommand{\calE}{{\mathcal{E}}}
\newcommand{\calF}{{\mathcal{F}}}

\newcommand{\calM}{{\mathcal M}}

\newcommand{\calP}{{\mathcal{P}}}

\newcommand{\hbP}{\widehat{\bP}} 

\newcommand{\hbTe}{\widehat{\bTe}}
\newcommand{\hbD}{\widehat{\bD}}
\newcommand{\hbU}{\widehat{\bU}}

\newcommand{\hr}{\widehat{r}} 
\newcommand{\hlam}{\widehat{\lam}}
\newcommand{\hK}{\widehat{K}}

\newcommand{\tilr}{\tilde{r}}
\newcommand{\tilR}{\tilde{R}} 
\newcommand{\tilbP}{\tilde{\bP}}
\newcommand{\tilbU}{\tilde{\bU}}

\long\def\ignore#1{}

\newcommand{\reals}{\mathbb{R}}

\title{\Large{\bf  Spectral clustering in the dynamic stochastic block model  }}

\author{
\large{ Marianna Pensky and Teng Zhang}  
  \\ \\
Department of Mathematics, University of Central Florida   } 

\date{\vspace{-5ex}}

\begin{document}

\maketitle
\begin{abstract}
In the present  paper, we studied a Dynamic Stochastic Block Model (DSBM) under the assumptions 
that the connection probabilities, as functions of time, 
are smooth and that  at most $s$ nodes can switch their class memberships between two consecutive time points.  
We estimate the edge probability tensor by a  kernel-type procedure and extract the group memberships of the nodes 
by spectral clustering. The procedure is computationally viable, adaptive to the unknown smoothness of the functional 
connection probabilities, to the rate $s$ of membership switching and to the unknown number of clusters.  
In addition, it is accompanied by non-asymptotic guarantees for the precision of estimation and clustering.

\vspace{2mm} 


\vspace{2mm}{\bf AMS (2000) Subject Classification}:  Primary: 62F12. Secondary: 62H30, 05C80.

\end{abstract} 


\section{Introduction  }
\label{sec:introduction}
\setcounter{equation}{0}

Networks arise in many areas of research such as sociology, biology, genetics, ecology, information technology
to list a few. An overview of statistical modeling of random graphs can be found in, e.g., 
 \cite{Kolaczyk:2009:SAN:1593430} and \cite{goldenberg2010survey}.
Analysis of stochastic networks   is extremely important and is used in a variety of applications 
such as sociology, biology, genetics, ecology, information technology and national security.
Stochastic networks are used, for example, to model brain connectivity,   
gene regulatory networks,    protein signaling networks, 
to monitor cyber and homeland security  and to evaluate and predict 
social relationships within   groups  or between groups such as  countries.

In this paper, we consider a dynamic network defined as an undirected graph with $n$ nodes
with connection probabilities changing in time. Specifically,  we observe adjacency matrices   $\bA_t$
 of the graph at  time instances  $\tau_t$ where $0 < \tau_1 < \cdots < \tau_T = b$.
Here, $\bA_t (i,j)$  are Bernoulli random variables with $\bP_t (i,j) = \Pr(\bA_t (i,j)=1)$ that are independent 
for any $1 \leq i < j \leq n$ and  $\bA_t (i,j)=\bA_t (j,i) = 1$ if a connection between nodes $i$ and $j$ is observed 
at time $\tau_t$ and  $\bA_t (i,j)=\bA_t (j,i)  = 0$  otherwise. We set $\bA_t(i,i)=10$  (or any other large enough constant)
and assume, for simplicity  that time instances are equispaced and the time interval is scaled to one, i.e. $b=1$ and $\tau_t = t/T$.

Furthermore, we assume that the network can be described by a Dynamic Stochastic Block Model (DSBM): 
at each time instant $\tau_t$   the nodes are grouped into $K$ classes   $G_{t,1}, \cdots, G_{t,K}$,
and the probability of a connection $\bP_t (i,j)$ is entirely  determined by the  groups to which the nodes $i$ and $j$ 
belong at the moment $\tau_t$. In particular,  if $i \in G_{t,k}$ and $j \in G_{t,k'}$, then 
$\bP_t (i,j) = \bB_t (k,k')$ where $\bB_t$ is the {\it connectivity matrix} at time $\tau_t$  with $\bB_t (k,k') = \bB_t (k',k)$. 
In this case, for any $t= 1, \ldots, T$, one has
\be \label{maineq}
\bP_t = \bTe_t \bB_t \bTe_t^T
\ee
where $\bTe_t$ is a {\it clustering} matrix  such that $\bTe_t$ has exactly one 1 per row
and $\bTe_t (i,k) =1$ if and only if    node $i$ belongs to the class $G_{t,k}$ and is zero otherwise.
One of the main problems in this setting is to cluster the nodes and identify the groups that 
have common probabilities of connections. If one had an oracle that would give the membership
assignments (matrices $\bTe_t$), then one could obtain accurate estimators of matrices $\bB_t$ 
and $\bP_t$ by averaging elements of the adjacency matrices.

The objective of the present paper is to suggest a modification of a popular spectral clustering procedure
to the case of the DSBM and study its precision at a time instant $\tau_t$ in a non-asymptotic setting.

The DSBM can be viewed as a natural extension of a Stochastic Block Model (SBM) 
which was extensively studies in the last decade (see, e.g., \cite{arias-castro2014},
\cite{bickel2009nonparametric}, \cite{choi2014co}, \cite{gao2015}, \cite{gao2015achieving},
\cite{jin2015fast},   \cite{joseph2016},  \cite{klopp2015oracle}, \cite{lei2015},
\cite{rohe2011},  \cite{sarkar2015}, \cite{verzelen2015},
 \cite{zhao2011community},  \cite{zhao2012} among others).

In comparison, there are many fewer results concerning the DSBM model.
Although   approaches developed for  time-independent networks can be  applied to a temporal network frame-by-frame,
they  totally ignore temporal continuity of the network  structures and parameters. 
Nonetheless, by taking advantage of continuity and observations at multiple time instances  
one can gain a better insight into a variety of issues and improve precision of the inference 
 (see \cite{han2015consistent}  and   \cite{pensky2016dynamic}).


A survey of the papers published before 2010  can be found in 
Goldenberg \etal \cite{goldenberg2010survey}.
After Olhede and Wolfe \cite{Olhede14102014} established universality of the SBM, 
several authors investigated the SBM in the dynamic setting. Majority of them  
described changes in the memberships via   Markov-type structures that allow  to 
model smooth evolution of groups across times. For example, Yang \etal \cite{yang2011detecting} 
assumed  that, for each node, its membership forms a Markov chain independent
of other nodes, however,  connection probabilities do not change in time.
 Xu and Hero III \cite{xu2014dynamic} allowed both the connection probabilities and the group memberships to
change with time via a   latent state-space model. Later, Xu \cite{xu2015stochastic} and Xu \etal \cite{Xu2014} further refined  the model 
by introducing a Markov structure on the memberships.  In both papers, the logits of connection 
probabilities are modeled  via a dynamical system model. Some authors  \cite{Herlau:2013:MTE:3042817.3043044}, \cite{yang2011detecting} 
presented Bayesian variants of similar ideas. 
We should also mention   Fu \etal \cite{Fu:2009:DMM:1553374.1553416} and Xing \etal \cite{xing2010} that extended   
the DSBM to the case of the mixed memberships under the assumption that data follows the multivariate logistic-normal distribution.
For example,  Xing \etal \cite{xing2010} assumed that the data followed  dynamic logistic-normal 
mixed membership blockmodel and inferred parameter values by using Laplace variational approximation
scheme which generalizes the variational approximation developed in Airoldi \etal \cite{Airoldi:2008:MMS:1390681.1442798}.
None of the   papers cited above inferred the number of classes. This shortcoming was corrected   
by   Matias and  Miele \cite{matias2016statistical} who propose  a 
Markov chain model for the membership transitions and infer the unknown parameters including the unknown 
number of classes via variational approximations of the EM algorithm. 
The  approach of \cite{matias2016statistical} was further extended by a very recent paper of 
Zhang \etal \cite{DBLP:journals/corr/ZhangMN16}
who assumed the   Poisson model on the  number of connections with the time-independent probabilities 
that edges appear or disappear at any time instant.   
We should also cite an early work of Chi \etal \cite{evolutionary-spectral-clustering-incorporating-temporal-smoothness} 
that made no assumptions on the mechanism that governs changes in the cluster memberships and deals 
with a problem by introducing two cost functions,
the snapshot cost associated with the error of current clustering and the temporal cost that measures how
the clustering preserves continuity in terms of cluster memberships where both  cost functions 
are based on the results of the $k$-mean clustering algorithm.


While some of the procedures described in those papers show good computational properties, 
they come without any guarantees for the accuracy of  estimation and clustering. To the best of our knowledge, 
the only paper that   investigates precision of temporal clustering is  \cite{han2015consistent},
where the authors apply the spectral clustering   to the  matrix   of averages
under the assumption that the  sequences  $\bB_t (k,k'), t=1, \ldots, T$ form   stationary ergodic processes
 for each $k$ and $k'$ and prove consistency of the procedure as $T$ and $n$   tend to infinity.

In this paper, we likewise consider a dynamic network that possesses  some kind of continuity
in a sense that neither connection probabilities $\bB_t$  in \fr{maineq} nor class memberships   change drastically
from one time instant to another. In particular, we assume that, for any pair  $k$ and $k'$ of classes, 
the connection probabilities $\bB_t (k,k')$  represent values of some smooth function at time $\tau_t$
and, therefore, can be treated as functional data. In addition, we suppose that at most $s$ nodes can switch 
their class memberships between two consecutive time points. Both assumptions guarantee some degree of consistency
of the network in time. Under those assumptions, we extract group memberships of the nodes at every time point 
by using  a spectral clustering procedure and evaluate the error of this procedure. 
The clustering technique is applied to   kernel-type   estimators of the edge 
probability matrices $\bP_t$ that we construct in the paper. By using Lepskii method, we achieve adaptivity of the suggested procedure
to the unknown temporal smoothness of the functional connection probabilities $\bB_t (k,k')$ and to the rate $s$ of membership
switching.  Finally, by setting a threshold on the ratio of the eigenvalues of the estimated probability matrix, we find $\hat{K}$, the estimated
number of clusters, that coincides with the true number of clusters $K$ with high probability.

Our paper makes several key contributions. We present a computationally viable methodology 
for estimating an edge probability matrix and clustering of a time-dependent  network
that follows the DSBM. The procedure is adaptive to the set of unknown parameters and  
is accompanied by non-asymptotic guarantees for the precision of estimation and clustering. 
In order to obtain those results, we develop a variety of new mathematical techniques. 
In particular, we develop a discrete kernel estimator for an unknown matrix and obtain 
its adaptive version using Lepskii method. To the best of our knowledge, neither of those 
methods have been used in this setting so far. In addition, 
our  analysis in Lemma~\ref{lem:variance} generalizes the methodologies 
developed by Friedman et al.~\cite{Friedman:1989:SER:73007.73063},  Feige and Ofek \cite{RSA20089}, 
and Lei and  Rinaldo \cite{lei2015}, from estimating the spectral norm of a random matrix to the spectral norm 
of the sum of independent random matrices. Our approach gives a sharper bound than the conventional matrix concentration 
inequalities such as the matrix Bernstein inequality~\cite{MAL-048}, by a logarithmic factor. 
Finally, we estimate the number of clusters and provide guarantees of the accuracy of this estimator.

The rest of the paper is organized as follows. Section~\ref{sec:notations} introduces notations 
and main assumptions of the paper. Section~\ref{sec:spec_clust} presents the spectral clustering 
algorithm and evaluates its error at time $t$ in terms of the estimation error of the matrix of 
the connection probabilities  $\bP_t$. For this reason, Section~\ref{sec:bPt_est} describes construction of  a kernel-type 
estimator of the probability matrix at each time point $t$ and evaluates its error. While 
Sections~\ref{sec:est_edgeprob}  and \ref{sec:est_error} assume  that the degree of smoothness $\beta$ of the 
connection probabilities   and the rate $s$ of switching of nodes' memberships  are known, Section~\ref{sec:adaptive}
utilizes the Lepskii method for construction of adaptive estimators of the connection probability matrices $\bP_t$.
Finally, Section~\ref{sec:clust_err} presents upper bounds for the clustering errors.
Section~\ref{sec:clust_num_est} offers an estimator for the number of clusters and provide precision guarantees 
for the clustering  procedure with the estimated number of clusters. 
Section~\ref{sec:discussion}   concludes the paper with a discussion.
Section \ref{sec:appendix}, Appendix, describes   construction of a discrete 
kernel and also contains proofs of all statements in the paper.


\section{Notations and assumptions } 
\label{sec:notations}
\setcounter{equation}{0}

For any $a,b \in \RR$, denote $a \vee b = \max(a,b)$, $a \wedge b = \min(a,b)$.
For any two positive sequences $\{ a_n\}$ and $\{ b_n\}$, $a_n \asymp b_n$ means that 
there exists a constant $C>0$ independent of $n$ such that $C^{-1} a_n \leq b_n \leq C a_n$
for any $n$. For any set $\Om$, denote cardinality of $\Om$ by $|\Om|$.
For any $x$, $[x]$ is the largest integer no larger than $x$.

For any vector $\bt \in \RR^p$, denote  its $\ell_2$, $\ell_1$, $\ell_0$ and $\ell_\infty$ norms by, 
respectively,  $\| \bt\|$, $\| \bt\|_1$,  $\| \bt\|_0$ and $\| \bt\|_\infty$.
Denote by $\bone$ and $\bzero$ vectors that have, respectively, only unit or zero elements.
Denote by $\boe_j$ the vector with 1 in the $j$-th position and all other elements equal to zero.

For a matrix $\bQ$, its $i$-th row and $j$-th columns are denoted, respectively, by
$\bQ_{i, *}$ and $\bQ_{*, j}$. Similarly,  reductions  of $\bQ$ to a set of rows or columns 
in a set $G$ are  denoted, respectively, by $\bQ_{G, *}$ and $\bQ_{*, G}$.
For any matrix $\bQ$,  denote its spectral and Frobenius norms by, respectively,  $\| \bQ \|$ and $\| \bQ \|_F$,
Denote the largest  in absolute value  element of $\bQ$ by $\| \bQ\|_{\infty}$
and the number of nonzero elements of $\bQ$ by $\|\bQ\|_0$.

Denote by $\calM_{n,K}$ the  collection of  clustering  matrices $\bTe \in \{0,1\}^{n\times K} $.
Denote by $n_t (k) = |G_{t,k}|$ the number of elements in class $G_{t,k}$  and let
$n_{t,\max} = \max_k n_t(k)$ and $n_{t,\min} = \min_k n_t(k)$, $k = 1, \ldots, K$.
We assume that there exists $\al_n$ independent of $T$ and an absolute constant $C_{\al}$
independent of $n$ and $T$ such that 
\be \label{al_ineq}
C_{\al}^{-1} \al_n \leq \| \bB_t\|_\infty \leq C_{\al}\, \al_n, \quad 1 \leq C_{\al} < \infty.
\ee
If the network is sparse, then $\al_n$ is small for large $n$ and   $\|\bP_t\|_\infty \leq C_{\al}\, \al_n$, 
otherwise, one can just set $\al_n=1$. Denote
\bes 
\bH_t = \al_n^{-1}\, \bB_t, \quad \bB_t = \al_n\, \bH_t.
\ees

We shall carry out time-dependent clustering of the nodes in the situation where 
neither the connection probabilities nor the cluster memberships change drastically 
from one time point to another. In addition, to make successful  clustering possible, the values 
of probabilities of connection should be sufficiently different from each other, which is guaranteed by the 
smallest eigenvalues of matrices $\bH_t$ being separated from zero.

In order to quantify those notions, we consider a H\"{o}lder class $\Sig(\beta, L)$  of functions $f$ on $[0,1]$
such that $f$ are $l$ times differentiable and 
\be \label{eq:holder}
|f^{(l)} (x) - f^{(l)} (x')| \leq L |x-x'|^{\beta -l}\quad \mbox{for any}\quad x,x' \in [0,1],
\ee
where $l$ is the largest integer strictly smaller than $\beta$. We suppose that the following assumptions hold.
\\

{\bf (A1). } For any $1 \leq k \leq k' \leq K$, there exist a function $f(\cdot;k,k')$ such that 
$\bH_t (k,k') = f(t/T;k,k')$ and $f(\cdot;k,k') \in \Sig(\beta, L)$.
\\

{\bf (A2). }  At most $s$ nodes can change their memberships between any consecutive time instances.
\\

{\bf (A3). }  There exists an absolute constant $C_{\lam}$, $1 \leq C_{\lam} < \infty$, 
independent of $n$ and $T$ such that
\bes
C_{\lam}^{-1} \leq \lam_{\min} (\bH_t) \leq \lam_{\max} (\bH_t) \leq C_{\lam}.
\ees

 Clustering of the nodes can be recovered only up to column permutations. 
However, in order condition {\bf A1} can hold, we shall assume that 
the node's labels are fixed and do not depend on $t$. We denote the set
of $K \times K$ permutation matrices by $\calE_K$ and, following \cite{lei2015},
consider two measures of clustering precision at time $\tau_t$. The first is
the overall relative clustering error at time $\tau_t$
\be \label{eq:Rt}
R_t (\hbTe_t, \bTe_t) = n^{-1}\ \min_{\bJ \in \calE_K} \|\hbTe_t \bJ - \bTe_t \|_0
\ee 
that measures the overall proportion of mis-clustered nodes.
The second measure is the highest relative  clustering error over the communities at time $\tau_t$
\be  \label{eq:tilRt}
\tilR_t (\hbTe_t, \bTe_t) =  \min_{\bJ \in \calE_K}\ \max_{1 \leq k \leq K}\ n_k^{-1}\ 
\|(\hbTe_t \bJ - \bTe_t)_{G_{t,k},*} \|_0. 
\ee 
In addition, we study two  global measures of clustering accuracy such as
the overall highest relative error over the communities and the  overall highest  relative error 
\be \label{high_clust_errors}
\tilR_{\max} = \max_{1 \leq t \leq T}\,   \tilR_t(\hbTe_t, \bTe_t), \quad
R_{\max} = \max_{1 \leq t \leq T}\,  R_t(\hbTe_t, \bTe_t)
\ee


\section{Spectral clustering and its error} 
\label{sec:spec_clust}
\setcounter{equation}{0}

Spectral clustering is a common method for community recoveries (see, e.g.,
\cite{jin2015fast}, \cite{joseph2016}, \cite{lei2015}, \cite{Luxburg:2007:TSC:1288822.1288832},
\cite{rohe2011}     and \cite{sarkar2015} among others). 
%
The accuracy of spectral clustering depends on how well 
one can relate the eigenvectors of $\bP_t = \bTe_t \bB_t \bTe_t^T$ to the eigenvectors of its estimator $\hbP_t$.
For this reason, our first goal will be to obtain an estimator $\hbP_t$ of $\bP_t$. 
Subsequently we shall apply the spectral clustering based on the approximate $k$-means algorithm  suggested by  
Lei and Rinaldo  \cite{lei2015}. Although one can read a description of the algorithm in their paper, for completeness
we review it here. \\

Given a matrix $\bP\in\mathbb{R}^{n\times n}$, let $\bU\in\mathbb{R}^{n\times K}$ be the matrix 
that consists of the first $K$ eigenvectors of $\bP$. Then  \cite{lei2015} 
suggested to investigate the $(1+\epsilon)-$approximate solution to the $k$-means problem 
applied to the $n$ row vectors of $\bU$, specifically, finding  $\hat{\bTe}\in\calM_{n,K}$ and  
$\hat{\bX}\in\mathbb{R}^{K\times K}$ that satisfy
\begin{equation}\label{eq:approximate}
\|\hat{\bTe}\hat{\bX}-\bU\|_F^2\leq (1+\epsilon)\ \min_{\stackrel{\bTe\in\calM_{n,K}}{\bX\in\mathbb{R}^{K\times K}}}\ 
\|\bTe\bX-\bU\|_F^2.
\end{equation}
Then the cluster assignments are   given by the estimated $\hat{\bTe}_t$. 
There exist efficient algorithms for solving \eqref{eq:approximate}, see, e.g., \cite{1366265}. 
The procedure is summarized as Algorithm 1.
\begin{algorithm}
\caption{\ Spectral clustering in the dynamic stochastic block model}
\label{alg:spectral}
{\bf Input:}  Adjacency matrices $\bA_t$ for $t=1,\ldots, T$; number of communities $K$;\\ 
\hspace*{12mm} approximation parameter $\epsilon$.  \\
{\bf Output:} Membership matrices $\bTe_t$ for any $t=1,\ldots, T$.\\
{\bf Steps:}\\
{\bf 1:} Estimate $\bP_t$ by $\hbP_{t,r}$ defined in  \eqref{eq:hbPt}. \\
{\bf 2:} Let $\bU_t\in\reals^{n\times K}$ be a matrix representing the first $K$ eigenvectors of $\hbP_{t,r}$.\\
{\bf 3:} Apply the $(1+\epsilon)$-approximate $k$-means algorithm to the row vectors of $\bU_t$ \\
{\bf 4:} Obtain   the solution  $\bTe_t$.
\end{algorithm}

The clustering errors $R_t(\hbTe_t, \bTe_t)$ in \fr{eq:Rt} and $\tilR_t(\hbTe_t, \bTe_t)$ in \fr{eq:tilRt} 
are  determined by the precision of estimation of $\bP_t$ by $\hbP_t$.
In particular, the following statement   holds.

\begin{lemma} \label{lem:clus_er_LR}
Let clustering be carried out according to the Algorithm 1 on the basis of an 
estimator $\hbP_{t,r}$ of $\bP_t$.
Let  $\bTe_t \in \calM_{n,K}$. Then,
\be \label{tilRtexp} 
\tilR_t(\hbTe_t, \bTe_t)   \leq   \frac{64 (2 + \eps) K}{\lam^2_{\min}(\bP_t)}\ \|\hbP_{t,r} - \bP_t\|^2 
\ee  
and, if the right-hand side of \fr{tilRtexp} is bounded by one, then
\be  \label{Rtexp}
R_t(\hbTe_t, \bTe_t)   \leq   \frac{64 (2 + \eps) K}{\lam^2_{\min}(\bP_t)}\ \frac{n_{t,\max}}{n}\ \|\hbP_{t,r} - \bP_t\|^2.
\ee 
Here, $\lam_{\min}(\bP_t)$ is the smallest nonzero eigenvalue of $\bP_t$.
\end{lemma}


\section{Estimation of the edge probability matrices}
\label{sec:bPt_est}
\setcounter{equation}{0}

\subsection{Construction of the estimator}
\label{sec:est_edgeprob}

In order to estimate $\bP_t$, we choose an integer $r \geq 0$, the width of the window, and consider 
three pairs of sets of integers
\beqns
\calF_{r,1} & =   \{-r, \cdots, r\},\  &\calD_{r,1}    = \{1+r, \cdots, T-r\};\\
\quad
\calF_{r,2} & =   \{0, \cdots, r\},\  &\calD_{r,2}    = \{1, \cdots, r \};\\
\quad
\calF_{r,3} & =   \{-r, \cdots, 0\},\  &\calD_{r,3}   = \{T-r+1, \cdots, T\}.
\eeqns
If $t \in \calD_{r,j}$, we construct an estimator of $\bP_t$ on the basis of $\bA_{t+i}$
where $i \in \calF_{r,j}$, $j=1,2,3$.  For this purpose, we introduce  discrete kernel functions $W^{(j)}_{r,l}(i)$
of an integer argument $i$   that satisfy the following assumption
\\

{\bf (A4). }  Functions $W^{(j)}_{r,l}$, $j=1,2,3$, are such that $|W^{(j)}_{r,l} (i)| \leq W_{\max}$, 
where $W_{\max}$ is independent of $r$, $j$ and $i$, and for $j=1,2,3$, one has
\be \label{eq:kerW}
\frac{1}{|\calF_{r,j}|}\ \sum_{i\in \calF_{r,j}}  i^k\, W^{(j)}_{r,l} (i) = \lfi
\begin{array}{ll} 
1, & \mbox{if}\ k=0, \\
0, & \mbox{if}\ k=1, \ldots, l.
\end{array} \right.
\ee 
Here $|\calF_{r,j}|$ is the cardinality of the set $\calF_{r,j}$.
\\

\noindent
One can easily see that function $W^{(j)}_{r,l}$ are discrete versions of   order $l$ continuous kernels
where $W^{(1)}_{r,l}$ corresponds to a regular kernel designed for the internal points of the interval
while $W^{(j)}_{r,l}$, j=2,3, mimic the boundary kernels (the left boundary kernel for $j=2$ and the right 
 boundary kernel for $j=3$).  Section \ref{sec:W_construct}   provides an algorithm 
for the explicit construction of $W^{(j)}_{r,l}$ for any values of $r,l$ and $j$.  We ought to point out that 
the dependence of $W^{(j)}_{r,l}$ on $r$ is a weak one, especially as $r$ grows. 
We  estimate the edge probability matrix $\bP_t$  by
\be \label{eq:hbPt}
\hbP_{t,r} = \sum_{j=1}^3 \ \II(t \in \calD_{r,j}) \lfi \frac{1}{|\calF_{r,j}|}\ \sum_{i\in \calF_{r,j}} W^{(j)}_{r,l} (i) \bA_{t+i} \rfi.
\ee
Note that since the sets $\calD_{r,j}$ are disjoint for different values of $j$, the estimator of $\bP_t$ always involve just one 
expression in figure brackets in formula \fr{eq:hbPt}.
\\


\subsection{Estimation error}
\label{sec:est_error}

In order to figure out how to choose the value of $r$, we evaluate the error $\|\hbP_{t,r} - \bP_t\|$.
Denote
\bes
\bP_{t,r} = \sum_{j=1}^3 \ \II(t \in \calD_{r,j}) \lfi \frac{1}{|\calF_{r,j}|}\ 
\sum_{i\in \calF_{r,j}} W^{(j)}_{r,l} (i) \bP_{t+i} \rfi 
\ees
and observe that 
\be \label{eq:Delt}
\Del_t (r) = \|\hbP_{t,r} - \bP_t\| \leq  \|\hbP_{t,r} - \bP_{t,r}\|+ \|\bP_{t,r}-   \bP_t\|
\equiv \Del_{1,t} (r) + \Del_{2,t} (r),
\ee
where $\Del_{1,t}  (r) =  \|\hbP_{t,r} - \bP_{t,r}\|$ and $\Del_{2,t}  (r) = \|\bP_{t,r}-   \bP_t\|$
represent, respectively, the variance and the bias portions of the error.
The following statements provide upper bounds for those quantities.

\begin{lemma} \label{lem:variance}
Let \fr{al_ineq} and Assumption {\bf A4} hold and $\alpha_n\geq   C_{\al}^{-1}\, c_0\,  \log n/n$.
For any $\tau>0$  there exists a set $\Om_{t,\tau}$ and a constant $C_{0,\tau} = C(\tau, c_0, C_{\al}, W_{\max})$ such that
$\Pr(\Om_{t,\tau}) \geq 1- 4\, n^{-\tau}$ and, for any $\om \in \Om_{t,\tau}$, one has 
\be \label{eq:lemvar}
\|\hbP_{t,r} - \bP_{t,r}\|\leq   C_{0,\tau}\,  \sqrt{n\, \al_{n}/(r \vee 1)}.
\ee
The exact expression for $C_{0,\tau}$ is given by formula \fr{eq:C0tau} in the Appendix.
\end{lemma}

\begin{lemma} \label{lem:bias}
Under Assumptions {\bf A1--A4} and \fr{al_ineq}, one has
\be \label{eq:bias}
 \|\bP_{t,r}-   \bP_t\| \leq \frac{L}{l!}\,  W_{\max}\, \al_n n\, \lkr \frac{r}{T} \rkr^\beta + 
2 \sqrt{2}\, W_{\max}\, C_{\lam}\, \al_n \sqrt{ n_{\max}\, r s}.   
\ee 
\end{lemma}

Lemmas \ref{lem:variance} and \ref{lem:bias} together with inequality \fr{eq:Delt} provide an upper bound
for  $\Del_t  (r)$. 
Since $\Del_{1,t}  (r)$ is decreasing and $\Del_{2,t}  (r)$ is increasing in $r$,
there exist a value $r^*$ that ensures the best bias-variance balance. Denote 
\be \label{eq:rstar}
r^* = \underset{r}{\operatorname{argmin}}\, \lkr  \|\bP_{t,r}-   \bP_t\| + C_{0,\tau}\,  \sqrt{n\, \al_{n}/(r \vee 1)} \rkr, 
\ee
\be \label{opt_er}
\del_1= \sqrt{n\,\al_{n}},\   \ 
\del_2 = \lkr \frac{\al_n^{\beta +1} n^{\beta +1}}{T^\beta} \rkr^{\frac{1}{2 \beta+1}} +
(\al_n^3\, n\, n_{\max}\, s)^{\frac{1}{4}},
\ee
Then, the following lemma provides an upper bound for $\Del_t (r)$.

\begin{lemma} \label{lem:opt_error}
Let \fr{al_ineq} and Assumptions {\bf A1--A4}   hold with  $\alpha_n\geq   C_{\al}^{-1}\, c_0\,  \log n/n$.
Then, the optimal value of $r$ is 
\be \label{r_opt}
r^* =    \min \lkr \lkv C_T\, \lkr n^{-1} \al_n T^{2\beta}\rkr^{1/(2 \beta+1)}\rkv,\ 
  \lkv C_s\, \sqrt{ (\al_n n_{\max}\, s)^{-1}\, n}\rkv  \rkr 
\ee
where $[x]$ is the largest integer no greater than $x$ and 
$C_T$ and $C_s$ are positive constants independent of $n$, $T$, $n_{\max}$, $s$ and $\alpha_n$. 
Also, with probability at least $1  - 4\, n^{-\tau}$ one has 
\be \label{eq:err_bound}
 \Del_t (r^*) \leq  C_{\Del}\ \min (\del_1, \del_2), 
\ee
where constant $C_{\Del}$ depends on $\tau, c_0, W_{\max}, \beta, L, C_{\al}$ and $C_{\lam}$.
\end{lemma}

Note that $\del_1 < \del_2$ corresponds to the case where $r^* =0$ and this situation occurs only 
if $T$ is rather small or $s$ is large. In particular, $\del_2 \leq \del_1$ if 
\be \label{eq:del1to2}
T \geq \tilde{C}_T (n/\al_n)^{\frac{1}{2\beta}}\quad  \mbox{and} \quad
s \leq \tilde{C}_s (n_{\max} \al_n)^{-1}\, n,
\ee 
where $\tilde{C}_T$ and $\tilde{C}_s$ are positive constants independent of $n$, $T$, $n_{\max}$, $s$ and $\alpha_n$. 
In this case, $r^* \geq 1$ and one can take an advantage
of the smoothness of the connection probabilities and the relative stability of group memberships.


\subsection{Adaptive estimation}
\label{sec:adaptive}

Observe  that the value of $r^*$ depends on the values of $s$, $n_{\max}$, $\al_n$ and $\beta$ that are  
unknown, therefore, in practice, the value $r^*$ in \fr{r_opt}  is unavailable. 
In order to construct an adaptive estimator we use the Lepskii method \cite{lepski1991}, \cite{lepski1997}.
For any $t$, set 
\be \label{rhat}
\hr \equiv \hr_t = \max \lfi 0 \leq r \leq T/2:\ \|\hbP_{t,r} - \hbP_{t,\rho}\| \leq 4\, C_{0,\tau}\, \sqrt{n\,\al_n/(\rho \vee 1)}\quad
\mbox{for any} \quad \rho < r \rfi
\ee
Observe that evaluation of $\hr$ does not require the knowledge of $s$, $n_{\max}$ or $\beta$.
If the network is not sparse,   one can set  $\al_n=1$.   Otherwise, one needs to know $\al_n$ for choosing an optimal
value of $r$. If $\al_n$ is known, the following lemma ensures that the replacement of $r^*$ by  $\hr$  changes the upper bound 
by a constant factor only.

\begin{lemma} \label{lem:adapt}
Let \fr{al_ineq} and Assumptions {\bf A1--A4}   hold with  $\alpha_n\geq   C_{\al}^{-1}\, c_0\,  \log n/n$.
Then, for any $\tau > 0$,  with probability at least $1 - 4\, n^{-\tau}$, one has   
\be \label{adap_err}
\|\hbP_{t,\hr} - \bP_t \| \leq 10 \, \min_r \lfi \|\bP_{t,r}-   \bP_t\| + C_{0,\tau}\,  \sqrt{n\, \al_{n}/(r \vee 1)}  \rfi
\leq 10 \, \Del_t (r^*).
\ee 
\end{lemma}


\section{The clustering error}
\label{sec:clust_err}
\setcounter{equation}{0}

Lemmas \ref{lem:clus_er_LR} and \ref{lem:adapt} allow  to obtain    upper bounds for the clustering errors.

\begin{theorem} \label{th:clus_er}
Let clustering be carried out according to the Algorithm 1.  
 Let $\bP_t = \bTe_t \bB_t \bTe_t^T$ where $\bB_t = \al_{n} \bH_t$  
and $\bTe_t \in \calM_{n,K}$. 
If \fr{al_ineq} and Assumptions {\bf A1--A4}   hold with $\alpha_n\geq   C_{\al}^{-1}\, c_0\,  \log n/n$, then  
for any $\tau > 0$,  with probability at least $1 -4\, n^{-\tau}$, one has   
\beqn
\tilR_t(\hbTe_t, \bTe_t) & \leq & C_R (2 + \eps)\,  \frac{K\, \min(\del_1^2, \del_2^2)}{\al_{n}^2\ n_{\min}^2}   
\label{tilRtexp1} \\
R_t(\hbTe_t, \bTe_t) & \leq & \tilR_t(\hbTe_t, \bTe_t) \ \frac{n_{\max}}{n},
\label{Rtexp1}  
\eeqn
where \fr{Rtexp1} holds provided the right-hand side of \fr{tilRtexp1} is bounded by one,
$\del_1$ and $\del_2$ are defined in \fr{opt_er} and $C_R = C_R(\tau, c_0, W_{\max}, \beta, L,  C_\al,  C_\lam)$.

In addition, if $T$ grows at most polynomialy with $n$, so that
$T \leq n^{\tau_1}$ for some $\tau_1 < \infty$, then   for any $\tau > 0$,  with probability at least 
$1 - 4\,  n^{-(\tau- \tau_1)}$, one has 
\be \label{eq:high_cl_er}
\tilR_{\max} \leq  C_R\,  (2 + \eps)   \,  \frac{K\, \min(\del_1^2, \del_2^2)}{\al_{n}^2\ n_{\min}^2},
\quad R_{\max} \leq \tilR_{\max}\ \frac{n_{\max}}{n}.
\ee  
\end{theorem}

Theorem \ref{th:clus_er}    provides upper bounds for the local clustering errors at 
time $\tau_t$ as well as for the maximum clustering errors on the  whole time interval. 
It confirms that as long as \fr{eq:del1to2} holds, at every time instance $\tau_t$,
the clustering errors obtained by our algorithm  will be smaller than those obtained by 
separately clustering snapshots of the network at each individual time point.
Consequently, the errors will be smaller than those reported in \cite{lei2015}.
In particular, by direct calculations, one derives 
\be \label{clust_err_explicit}
 \frac{K\, \min(\del_1^2, \del_2^2)}{\al_{n}^2\ n_{\min}^2} \leq \frac{K n}{\al_n\, n_{\min}^2}\, 
  \min \lkr 1; \lkr \frac{n \, \al_n}{T^{2\beta}} \rkr^{\frac{1}{2\beta+1}} + \sqrt{\frac{n_{\max}\, \al_n s}{n}}\, \rkr
\ee
For example, if the community sizes are   balanced, i.e. there exist positive constants $C_1$ and $C_2$
such that
\be \label{com_size}
C_1 \, \frac{n}{K} \leq n_{\min} \leq   n_{\max} \leq C_2 \, \frac{n}{K},
\ee
we immediately obtain the following corollary.

\begin{corollary} \label{cor1}
Under assumptions of Theorem \ref{th:clus_er}, for any $\tau > 0$,  with probability at least $1 - 4\, n^{-\tau}$, one has   
\be \label{balanced_err}
\tilR_t(\hbTe_t, \bTe_t)  \leq \tilde{C}_R (2 + \eps) \, \frac{K^3}{n \al_n}\,  
\min \lfi 1;\ \lkr \frac{n \, \al_n}{T^{2\beta}} \rkr^{\frac{1}{2\beta+1}} +
+ \sqrt{\frac{s \al _n}{K}} \rfi,   
\ee
where $\tilde{C}_R = \tilde{C}_R(\tau, c_0, W_{\max}, \beta, L,  C_\al,  C_\lam)$, and also \fr{Rtexp1} holds
provided the right hand side of \fr{balanced_err} is bounded by one.
Moreover, with probability at least $1 - 4\,  n^{-(\tau- \tau_1)}$, one has 
\bes  
\tilR_{\max} \leq  \tilde{C}_R (2 + \eps) \, \frac{K^3}{n \al_n}\,  
\min \lfi 1;\ \lkr \frac{n \, \al_n}{T^{2\beta}} \rkr^{\frac{1}{2\beta+1}} +
+ \sqrt{\frac{s \al _n}{K}} \rfi.
\ees
\end{corollary}

\begin{remark} \label{rem:dense_graph}
{\bf  Dense network. }{\rm
Inequalities \fr{tilRtexp1}, \fr{Rtexp1}, \fr{balanced_err}  and \fr{eq:high_cl_er}  
imply that precision of clustering is better when $\al_n$ is larger. Indeed, if the network is dense,
then $\al_{n} =1$, the estimator $\hbP_{t,\hr}$ is fully adaptive and  with probability at least 
$1 - 4\, n^{-(\tau- \tau_1)}$, 
$$
\tilR_{\max} \leq C_R \,  (2 + \eps)  \, 
 \frac{K\, n}{n_{\min}^2}\, 
  \min \lkr 1; \lkr \frac{n}{T^{2\beta}} \rkr^{\frac{1}{2\beta+1}} + \sqrt{\frac{n_{\max}\, s}{n}} \rkr.
$$
}
\end{remark}

\begin{remark} \label{rem:const_member}
{\bf Constant memberships. }{\rm
If  group memberships of the nodes remain unchanged over time,
then $s=0$ and one can cluster the average $\overline{\bP}$ of edge probability matrices 
on the basis of its observed counterpart  $\hbP$ where
\bes
\overline{\bP} = T^{-1} \ \sum_{t=1}^T \bP_t, \quad \hbP = T^{-1} \ \sum_{t=1}^T \bA_t
\ees
rather than the individual matrices $\bP_t$. In this case, 
$2r+1=T$, $W_{\max} =1$ and the bias portion of the error disappears,
hence, 
\be \label{eq:con_mem}
\|\hbP - \overline{\bP}\| \leq C_{0,\tau} \sqrt{n \al_n/T}.
\ee
Observe that $\lam_{\min}(\hbP) \geq C_\lam^{-1}\,  \al_{n} \, n_{\min}$.
Therefore,  for any $\tau > 0$,  with probability at least $1 - 4\, n^{-\tau}$, one has  
$$
\tilR_t(\hbTe_t, \bTe_t)  \equiv \tilR (\hbTe, \bTe) \leq   64 (2 + \eps)\,  C_{0 \tau}^2\, C_\lam^2  \,   
\frac{K\, n}{T\, \al_n n_{\min}^2}, \quad
R (\hbTe, \bTe) \leq \tilR (\hbTe, \bTe) \, \frac{n_{\max}}{n}. 
$$
}
\end{remark}

\begin{remark} \label{rem:const_probmatr}
{\bf Constant matrix of connection probabilities. }{\rm
Consider the situation where  nodes of the network can switch memberships in time ($s>0$)
but the matrix of the connection probabilities is constant:\ $\bB_t \equiv \bB$. In this case, Assumption {\bf A1}
is valid with $\beta = \infty$. Then, for $r < T$ one has 
$ \|\bP_{t,r}-   \bP_t\| \leq  2 \sqrt{2}\, W_{\max}\, C_{\lam}\, \al_n \sqrt{ n_{\max}\, r s}.$
Hence, $\del_2 = (\al_n^3\, n\, n_{\max}\, s)^{\frac{1}{4}}$  and
for any $\tau > 0$,  with probability at least $1 -4\, n^{-\tau}$, the clustering error at time $\tau_t$  appears as
\bes
\tilR_t(\hbTe_t, \bTe_t)  \leq   C_R (2 + \eps)\, \frac{K n}{\al_n\, n_{\min}^2}\, 
  \min \lkr 1;   \sqrt{\frac{n_{\max}\, \al_n s}{n}}\, \rkr
\ees
}
\end{remark}


\section{ Estimating the number of clusters} 
\label{sec:clust_num_est}
\setcounter{equation}{0}

Lemma~\ref{lem:adapt} implies that the eigenvalues of $\hbP_{t,\hr}$ can be used to 
estimate the true number of clusters $K$. Indeed, denote the sorted eigenvalues 
of any symmetric matrix $\bX\in\mathbb{R}^{n\times n}$ 
by $\lambda_1(\bX)\geq \lambda_2(\bX)\geq \ldots\geq \lambda_n(\bX)$.
Then, due to the  matrix perturbation result $|\lambda_i(\bX)-\lambda_i(\bY)|<\|\bX-\bY\|$
(see, e.g., Corollary III.2.2  of \cite{Bhatia1997}), obtain   
\bes
\lambda_{K+1}(\hbP_{t,\hr})\leq \|\hbP_{t,\hr} - \bP_t\|, 
\quad \lambda_{j}(\hbP_{t,\hr})\geq  \lambda_{j}(\bP_t) - \|\hbP_{t,\hr} - \bP_t\|, \ \ j=1, \cdots, K.
\ees
Denote $\lam_{j,t} = \lam_j(\bP_t)$, $\hlam_{j,t} = \lam_j(\hbP_{t,\hr})$   and  
\bes 
\eps_t = \Del_t(\hr_t)/\lam_{K,t} \quad \mbox{with} \quad \Del_t(\hr_t) = \|\hbP_{t,\hr} - \bP_t\|.
\ees
Then,  one has 
\be \label{eig_ratios}
\frac{\hlam_{j+1,t}}{\hlam_{j,t}} \geq \frac{1 - \eps_t}{\lam_{j,t}/\lam_{j+1, t} + \eps_t},\ j=1, \cdots, K-1,
\quad \quad
\frac{\hlam_{K+1,t}}{\hlam_{K,t}} \leq \frac{\eps_t}{1 - \eps_t}.
\ee 
Hence, if $\eps_t$ is small enough, then there exists a threshold $\varpi$ such that
\be \label{thresh_ineq}
\frac{\hlam_{j+1,t}}{\hlam_{j,t}} > \varpi, \ j=1, \cdots, K-1, \quad
\frac{\hlam_{K+1,t}}{\hlam_{K,t}} \leq \varpi
\ee 
while $\hlam_{j+1,t}/\hlam_{j,t}$ can exhibit chaotic behavior for $j \geq K+1$.
Therefore, one can estimate $K$ by   
\be \label{K_est}
\hK_t = \min \lfi k:\ \hlam_{k+1,t} < \varpi \hlam_{k,t} \rfi
\ee
where $\varpi$ is a tuning parameter. Note that the expression for $\hK_t$ is somewhat similar to 
the one suggested by Le and Levina \cite{2015arXiv150700827L} with the difference that we use 
eigenvalues of the adjacency  matrix  in the situation of a time-dependent network.

The following statement shows that if eigenvalues of $\bP_t$ grow at most exponentially and 
$\lam_{K,t} = \lam_{\min} (\bP_t)$  is large enough, then
$\hat{K}_t$ is an accurate estimator of $K$ with high probability.

\begin{proposition} \label{prop:clust_est}
Let Assumptions {\bf A1--A3} hold and $\alpha_n\geq   C_{\al}^{-1}\, c_0\,  \log n/n$. Let for some $w>0$
\be \label{eig_exp_cond}
\lam_{j} (\bP_t) \leq (1 + w)\, \lam_{j+1} (\bP_t),\ \ j=1, \cdots, K-1,
\ee
where $K$ is the true number of clusters. If 
\be \label{min_eig_con}
\lam_{\min} (\bP_t) \geq (40 + 10 w) \,  \Del_t (r^*)
\ee
where $ \Del_t (r^*)$ is defined in \fr{eq:err_bound}, then 
for any $\tau > 0$, with probability at least $1 - 4\, n^{-\tau}$, inequalities \fr{thresh_ineq} 
hold with $\varpi = (3+w)^{-1}$ and $\hK_t = K$.
\end{proposition}

Observe that condition \fr{min_eig_con} on the lowest nonzero eigenvalue of $\bP_t$ is essentially
a necessary condition required for   accurate clustering. Indeed, 
$\Del_t (r^*) \leq C_{\Del}\ \min (\del_1, \del_2)$ by \eqref{eq:err_bound} and, by Assumption {\bf A3}, 
$\lam_{\min} (\bP_t) \geq C_{\lam}^{-1}\, \al_n n_{\min}$, so that 
\eqref{min_eig_con} is guaranteed by 
\be \label{min_eig_con_guarantee}
\aleph = \frac{\min (\del_1^2, \del_2^2)}{\al_n^2 n_{\min}^2} \leq \tilde{C} = (C_{\Del} \, C_{\lam} \, (40 + 10 w))^{-2}. 
\ee
On the other hand, the clustering error in \eqref{tilRtexp1}  is bounded above by 
$\tilR_t(\hbTe_t, \bTe_t)   \leq   C_R (2 + \eps)\, K\, \aleph$ where $\aleph$ is defined in \fr{min_eig_con_guarantee}.
Therefore,  a small value $\tilR_t(\hbTe_t, \bTe_t) \leq \del$
of the clustering error implies that $\aleph \leq (C_R (2 + \eps))^{-1} \del/K$ which ensures \fr{min_eig_con_guarantee}
provided that $K$ is large enough.

Note also that assumption \fr{eig_exp_cond} is not restrictive. Indeed, since 
$\lam_K (\bP_t) = \lam_{\min} (\bP_t) \geq C_{\lam}^{-1}\, \al_n n_{\min}$ and 
$\lam_1 (\bP_t) = \lam_{\max} (\bP_t) \leq C_{\lam} \, \al_n n_{\max}$, obtain that
\bes
\frac{\lam_1 (\bP_t)}{\lam_K (\bP_t)} \leq C_{\lam}^2 \, \frac{n_{\max}}{n_{\min}},
\ees
so condition \fr{eig_exp_cond} always holds, for example,  in the case of a balanced model satisfying 
\fr{com_size}.

Combination of Theorem \ref{th:clus_er} and Proposition \ref{prop:clust_est} immediately
yield the following corollary.

\begin{corollary} \label{cor3}
Let clustering be carried out according to the Algorithm 1 with $\hK$ clusters.
Let assumptions of Theorem \ref{th:clus_er} be valid and \fr{eig_exp_cond} and \fr{min_eig_con} hold.
Then, for any $\tau > 0$,  with probability at least $1 - 4\, n^{-\tau}$,  
\fr{tilRtexp1} and \fr{Rtexp1} hold.
 \end{corollary}


\section{Discussion}
\label{sec:discussion}
\setcounter{equation}{0}

In the present  paper, we study the DSBM under the assumptions that the connection probabilities, as functions of time, 
are smooth and that  at most $s$ nodes can switch their class memberships between two consecutive time points.  
We estimate the edge probability tensor by a  kernel-type procedure and extract group memberships of the nodes 
by spectral clustering. The methodology is computationally viable, adaptive to the unknown smoothness of the functional 
connection probabilities, to the rate $s$ of membership switching and to the unknown number of classes.  
In addition, it is accompanied by non-asymptotic 
guarantees for the precision of both estimation and clustering.

Since we do not make an assumption that the network is generated by a time-varying graphon,
we cannot take advantage of the techniques described in \cite{2016arXiv160701718E} or
 \cite{2015arXiv150908588Z}.
Nevertheless, under an appropriate set of conditions, one can possibly improve the precision of clustering
using approaches developed in those papers.


\section*{Acknowledgments}

Marianna Pensky   was  partially supported by National Science Foundation
(NSF)  grant DMS-1407475.


\section{Appendix } 
\label{sec:appendix}
\setcounter{equation}{0}

\subsection{Construction of   kernels of integer arguments}
\label{sec:W_construct} 

First consider construction of the kernel $W^{(1)}_{r,l}$ designed for internal points.
Since it has symmetric domain, assume, without loss of generality that $l$ in \fr{eq:kerW} is even,
that is $l = 2m$. Consider $W^{(1)}_{r,l} (i)$ of the form
\bes 
W^{(1)}_{r,l} (i) = \sum_{j=0}^m a_j r^{-2j} i^{2j} 
\ees
where coefficients $a_j, j=0, 1 \ldots, m,$ are to be determined. In this case, \fr{eq:kerW}
is automatically valid for odd values of $k$. 
Denote 
\be \label{calP}
\calP_0 =1;\quad \calP_h (r) =  r^{-(h+1)} \, \sum_{i=1}^r i^{h},\ h= 0,2,\ldots. 
\ee  
Observe that $\calP_h (r)$ are polynomials in $1/r$ of degree $h+1$  and that expressions for $\calP_h (r)$,
 are known exactly for every $h=1, \cdots, m$ (see  e.g., \cite{gradshteyn2014table}, formula 0.121).
For example, $\calP_2 (r) =  (1 + r^{-1})(2 + r^{-1})/6$ and 
$\calP_4 (r) = r^{-4}\,(1+r^{-1})(2 + r^{-1})(3  + 3r^{-1} - r^{-2})/30$.
Then, conditions \fr{eq:kerW} can be re-written as 
\be \label{eq:ker1} 
r^{-1} \sum_{j=0}^m a_j\, r^{-2j} \sum_{i=-r}^r i^{k + 2j} = \II(k=0)
\ee 
and it holds for every odd value of $k$. If $k$ is even, $k = 2 k_0$, then
\fr{eq:ker1}  leads to the following system of $(m+1)$ linear equations in 
$a_j, j=0, \ldots, m$
\be  \label{eq:ker2}
a_0 (1 + 1/(2r)) + \sum_{j=1}^m a_j \calP_{2j} (r) =   1/(2r);\quad
\sum_{j=0}^m a_j \calP_{2(k_0+ j)} (r) = 0,\quad k_0  = 1,2, \cdots,  m.
\ee  
It is easy to check that the determinant of the system of equations \fr{eq:ker1}
is nonzero and, for every $r \geq 1$, the system \fr{eq:ker2} allows to recover a kernel  $W^{(1)}_{r,l}$.
For example, for $m=1$  we derive
\bes
a_0 = \frac{\calP_4(r)}{ \calP_4(r) - [\calP_2(r)]^2} = \frac{6(3  + 3r^{-1} - r^{-2})}{8  + 3r^{-1} -11 r^{-2}}; \quad 
a_1 = - \frac{\calP_2(r)}{\calP_4(r) - [\calP_2(r)]^2}= \frac{30}{11 r^{-2}  - 3r^{-1} - 8}.
\ees 
Construction of the boundary kernels $W^{(j)}_{r,l}$, $j=2,3$, are very similar to
$W^{(1)}_{r,l}$. For the sake of brevity, we describe only construction of $W^{(2)}_{r,l}$.
Write $W^{(2)}_{r,l}$ in a form
\bes 
W^{(2)}_{r,l} (i) = \sum_{j=0}^l b_j r^{-j} i^{j} 
\ees
and choose the coefficients, so that the kernel satisfies condition \fr{eq:kerW}.
The latter leads to the system of $(l+1)$ linear equations of the form
\bes 
b_0 (1 + r^{-1}) + \sum_{j=1}^l b_j \calP_j (r) = 1 + r^{-1}; \quad 
\sum_{j=0}^l b_j \calP_j (r) = 0.
\ees




\subsection{Proofs of the statements in the paper}
\label{sec:proofs}

\noindent
{\bf Proof of Lemma \ref{lem:clus_er_LR}. }
If $\bP_t = \bU \bD \bU^T$ and $\hbP_t  = \hbU \hbD \hbU^T$, then, by Lemma 5.1  of  \cite{lei2015}, 
obtain that there exists and orthogonal matrix $\bO$  such that 
\bes 
\|\hbU - \bU \bO \| \leq \frac{2 \sqrt{2 K}}{\lam_{\min} (\bP_t)} \, \|\hbP_t - \bP_t \|.
\ees
Let $S_{t,k}$ is a subset of nodes in class $G_{t,k}$ that are misclassified.
Then, Lemma~5.3 and Theorem~3.1 of Lei and Rinaldo (2015) imply that  
\be \label{eq:cl_er1}
\sum_{k=1}^K \frac{|S_{t,k}|}{n_t (k)} \leq 8 (2 + \eps) \, \|\hbU - \bU \bO \|^2
\leq \frac{64 K (2 + \eps)}{[\lam_{\min} (\bP_t)]^2}  \, \|\hbP_t - \bP_t \|^2
\ee  
provided the right-hand side of \fr{eq:cl_er1} is bounded by one.
In order to derive \fr{Rtexp}, observe that
\bes
R_t(\hbTe_t, \bTe_t) = \sum_{k=1}^K |S_{t,k}| \leq 
\frac{n_{t,\max}}{n} \,  \sum_{k=1}^K \frac{|S_{t,k}|}{n_{t}(k)}.
\ees
 \\

\medskip


\noindent
{\bf Proof of Lemma \ref{lem:variance}. }
Since the case $r=0$ follows directly from Theorem 1.1 in the Supplementary material of~\cite{lei2015}, 
we can assume that $r\geq 1$. Also, in order to simplify the proof, we do not consider kernels $W_{r,l}^{(j)}$ 
for each $j=1,2,3$, separately, but instead remove the index $j$ since the proofs are practically identical 
for all three values of $j$.

Lemma 2.1 in the Supplementary material of~\cite{lei2015} (with $\del = 1/2$) implies that
\begin{equation}\label{eq:deltanet}
\|\hbP_{t,r} - \bP_{t,r}\|\leq 4\sup_{\bx,\by\in\mathcal{T}}|\bx^T(\hbP_{t,r} - \bP_{t,r})\by|,
\end{equation}
where 
\begin{equation}
\mathcal{T}=\{\bx = (x_1,\cdots, x_n) \in \reals^n, \|\bx\|=1,\quad \text{$2\sqrt{n}x_i$ are all integers}.\}
\end{equation}
Hence,  we  bound above the right-hand side   of \eqref{eq:deltanet} by dividing the coordinates of $\bx$ and $\by$ 
into ``light pairs'' and ``heavy pairs'' as follows
\[
\mathcal{L}(\bx,\by)=\Big\{(i,j): |x_i y_j|\leq \sqrt{\frac{C_{\al} \al_n r}{n}} \Big\}, \quad 
\bar{\mathcal{L}}(\bx,\by)=\Big\{(i,j): |x_iy_j|> \sqrt{\frac{C_{\al} \al_n r}{n}}\Big\}. 
\]
Note that here the definition is different from the proof in~\cite{lei2015} by a factor of $r$, 
since we consider weighted sum of random matrices (instead of a single random matrix).
Partitioning $|\bx^T(\hbP_{t,r} - \bP_{t,r})\by|$ into the portions containing the ``light pairs'' and the ``heavy pairs'',
obtain
\begin{equation}  \label{eq:lightheavy}
\Big|\bx^T(\hbP_{t,r} - \bP_{t,r})\by\Big|
\leq  \Big|\sum_{(i,j)\in\mathcal{L}(\bx,\by)}x_i [\hbP_{t,r} - \bP_{t,r}](i,j) y_j\Big|
+ \Big|\sum_{(i,j)\in\bar{\mathcal{L}}(\bx,\by)}x_i[\hbP_{t,r} - \bP_{t,r}](i,j) y_i\Big|.
\end{equation}
In order to obtain upper bounds for the right-hand side of \fr{eq:lightheavy}, we need  
three supplementary statements, Lemmas \ref{lemma:lightpairs}, \ref{lemma:heavypairs1} and \ref{lemma:heavypairs2},
that generalize, respectively, Lemmas 3.1, 4.1 and 4.2 of Lei and Rinaldo  \cite{lei2015}
to our setting. The proofs of  Lemmas \ref{lemma:lightpairs}, \ref{lemma:heavypairs1} and \ref{lemma:heavypairs2}  
are deferred till Section \ref{sec:suppl_proofs}.

\begin{lemma}
\label{lemma:lightpairs}
Under assumptions of Lemma~\ref{lem:variance}, one has
\[
\Pr\lfi \sup_{\bx,\by\in\mathcal{T}}\Big|\sum_{(i,j)\in\mathcal{L}(\bx,\by)} x_i  y_j
[\hbP_{t,r} (i,j) - \bP_{t,r}(i,j)]  \Big| \leq C_{\tau,1} W_{\max}  \sqrt{\frac{C_{\al}\, n\, \al_n}{r}}\rfi
\geq 1 - 2 n^{-\tau} 
\]
provided
\be \label{Ctau1}
C_{\tau,1} \geq \max \lfi 2 \sqrt{(\tau + \log 14)},\ 8 (\tau + \log 14)/3 \rfi.
\ee
\end{lemma}

\medskip

\begin{lemma}
\label{lemma:heavypairs1}
Let $d_{t}(i)$ be the degree of the $i$-th node in the network with connection probabilities given by the matrix $\bP_t$. 
Denote
$$
d_{t,r}(i)=\sum_{k\in \calF_{r}} W_{r,l}(k) d_{t+k}(i).
$$ 
Then, under assumptions of Lemma~\ref{lem:variance}, one has
\be \label{eq:assumption_degree}
\Pr \lfi \max_{1 \leq i \leq n} d_{t,r}(i) \leq 3 C_{\al}\, (W_{\max}C_{\tau,2} +1)\, n  \al_n r \rfi \geq 1 - n^{-\tau}
\ee
provided
\be \label{Ctau2}
C_{\tau,2} \geq \max \left(\sqrt{\frac{2  (\tau +1)}{c_0}}, \frac{\tau+1}{3 c_0} \right).
\ee
\end{lemma}

\medskip

\begin{lemma}
\label{lemma:heavypairs2}
Let $I,J\subseteq \{1, \cdots, n\}$ with $|I|\leq |J|$. Denote   
$$
\bar{\mu}(I,J)= C_{\al} \alpha_n|I||J|\, |\calF_{r}|, \quad e_{t,r}(I,J)=\sum_{k\in \calF_{r}} W_{r,l}(k) e_{t+k}(I,J),
$$ 
where $e_{l}(I,J)$ represents the number of edges between $I$ and $J$ in the network at time $t+k$. 
Assume that the event in \eqref{eq:assumption_degree} holds. Then with probability at least $1-n^{-\tau}$, at least one of the following is true:
\begin{enumerate}
\item $e_{t,r}(I,J) \leq e\, C_{\tau,3} \, \bar{\mu}(I,J) $,
\item  $e_{t,r}(I,J)\, \log \lkr \frac{e_{t,r}(I,J)}{\bar{\mu}(I,J)} \rkr
\leq C_{\tau,4}\, |J|\, \log \lkr \frac{n}{|J|} \rkr$.
\end{enumerate}
Here 
\be \label{Ctau34}
C_{\tau,3}=\max\{3(W_{\max}C_{\tau,2} +1),e^{3W_{\max}}+1\}, \quad 
 C_{\tau,4}= 8\, W_{\max} (\tau +6).
\ee
\end{lemma}

The first term in the right-hand side of \eqref{eq:lightheavy} corresponding to  the ``light pairs''  is
bounded by  Lemma~\ref{lemma:lightpairs}. In order to bound the ``heavy pairs'' in the second term, observe that 
\begin{align} \label{eq:heavy1}
 &\left|\sum_{(i,j)\in\bar{\mathcal{L}}(\bx,\by)} x_i  y_j\, \bP_{t,r}(i,j) \right| \leq  
 \frac{1}{|\calF_{r}|}\  \sum_{(i,j)\in\bar{\mathcal{L}}(\bx,\by)}\ \sum_{k \in \calF_{r}}  
 \frac{x_i^2 y_j^2}{|x_i y_j|}\, \left|W_{r,l}(k)\right|\, \bP_{t+k}(i,j) \\
 \leq &\frac{1}{|\calF_{r}|}\sqrt{\frac{n}{C_{\al} \alpha_n r}}\sum_{ {k \in \calF_{r}} } 
\left|W_{r,l}(k)\right|\, C_{\al} \alpha_n \sum_{{ (i,j)\in\bar{\mathcal{L}}(\bx,\by)}} 
x_i^2 y_j^2 \leq 
W_{\max} \sqrt{\frac{C_{\al}\, n\, \alpha_n}{r}}.\nonumber
\end{align}

Applying Lemmas~\ref{lemma:heavypairs1}~and~\ref{lemma:heavypairs2},  and the same argument as in Section 4 
in the Supplementary material of \cite{lei2015} with (with $C_{\al}\, n\,\alpha_n\, r$ replacing $d$), we derive
\begin{align} \label{eq:heavy2}
\Pr \lfi \frac{1}{|\calF_{r}|}\ \left|\sum_{ \stackrel{ (i,j)\in\bar{\mathcal{L}}(\bx,\by) }{ k \in \calF_{r}} }
x_i \, y_j\, W_{r,l}(k)\,\bA_{t+k}(i,j) \right| \leq \tilde{C}_\tau \ \sqrt{ \frac{C_{\al} n\, \alpha_n}{r}} \rfi 
\geq 1- 2 n^{-\tau},
\end{align}
where $\tilde{C}_\tau =8\, \{16\delta^{-2 }+e\,C_{\tau,3}\delta^{-2}+ 24\, (W_{\max}C_{\tau,2}+1) +
 40\, C_{\tau,4} + 8\}$ with $\delta=1/2$. 
Combining \eqref{eq:heavy1} and \eqref{eq:heavy2}, obtain that 
the second term in the right-hand side  of \eqref{eq:lightheavy} is bounded above by 
$(W_{\max}+ \tilde{C}_\tau)\, \sqrt{C_{\al}\, n\, \alpha_n/r}$, with probability at least $1 - 2 n^{-\tau}$.

Combining   \eqref{eq:deltanet}, \eqref{eq:lightheavy}, Lemma~\ref{lemma:lightpairs},
\eqref{eq:heavy1} and \eqref{eq:heavy2},   we derive
\[
\Pr\left\{\|\hbP_{t,r} - \bP_{t,r}\|\leq 4(W_{\max}+W_{\max} C_{\tau,1} + \tilde{C}_\tau)\, \sqrt{\frac{C_{\al}\,  n\, \al_n}{r}}\right\}\geq 
1-4 n^{-\tau}, 
\]
and  obtain the expression for  $C_{0,\tau}$ in \eqref{eq:lemvar}:
\be \label{eq:C0tau}
C_{0,\tau} = 4 \sqrt{C_{\al}}\, \{ W_{\max} (1 +   C_{\tau,1}) +  
32 (24 \, W_{\max}C_{\tau,2} +  4 e\, C_{\tau,3} + 40\, C_{\tau,4} + 96) \}
\ee
where $C_{\tau,1}, C_{\tau,2}, C_{\tau,3}$ and $C_{\tau,4}$ are defined in \eqref{Ctau1},
\eqref{Ctau2} and \eqref{Ctau34}, respectively.
\\
 

\noindent
{\bf Proof of Lemma \ref{lem:bias}. }
First, let us prove that under Assumption {\bf A2}, one has for any $k$ such that $1 \leq t+ k \leq T$
one has
\be \label{eq:dif}
 \|\bTe_t - \bTe_{t+k}\| \leq \sqrt{2 |k| s}.
\ee
Let, without loss of generality, $k>0$.
Note that matrix $\bTe_t - \bTe_{t+k}$ at most $ks$ nonzero rows in which one entry is 1 and another is -1. 
If we permute the rows of matrix $\bTe_t - \bTe_{t+k}$ so that those nonzero rows are the first ones, 
we obtain that $(\bTe_t - \bTe_{t+k})(\bTe_t - \bTe_{t+k})^T$ is the block-diagonal matrix 
with the only nonzero block matrix $\bLam \in \RR^{ks \times ks}$ in the top left corner
that has elements with absolute values equal to 0,1 or 2. 
Then 
\bes
\lam_{\max} \lkv(\bTe_t - \bTe_{t+k})(\bTe_t - \bTe_{t+k})^T \rkv = 
\lam_{\max} (\bLam) \leq \max_{1 \leq i \leq ks} \sum_{i=1}^{ks} \leq 2 |k| s
\ees
which implies \fr{eq:dif}. Note that the upper bound is tight (to see this,
consider the case where s elements move from class $i$ to class $j$ at each 
of $k$ time points.

Next, note that $\Del_{2,t}$ in \fr{eq:Delt} can be decomposed as 
\be  \label{eq:Del2t}
\|\bP_{t,r}-   \bP_t\| \leq \|\bP_{t,r}- \tilbP_{t,r}\| + \| \tilbP_{t,r}- \bP_t\| 
\equiv \Del_{2,1,t} + \Del_{2,2,t},
\ee 
where 
\beqns
\bP_{t,r}    & = & r^{-1}\ \sum_{i=-r}^r W_{r,l} (i) \bP_{t+i}\\
\tilbP_{t,r} & = & r^{-1}\ \sum_{i=-r}^r W_{r,l} (i) \bTe_t \bB_{t+i} \bTe_t^T
\eeqns
Let us show that 
\be \label{eq:Del21t}
\Del_{2,1,t} = \|\bP_{t,r}- \tilbP_{t,r}\| \leq 
2 \sqrt{2}\, W_{\max}\, C_{\lam}\, \al_n \sqrt{ n_{\max}\, r s}
\ee
For this purpose observe that $\|\bTe_t\| \leq \sqrt{n_{\max}}$ and that 
\bes
\Del_{2,1,t} =  \| r^{-1}\ \sum_{i=-r}^r W(i)  \lkv \bTe_{t+i} \bB_{t+i} \bTe_{t+i}^T - \bTe_t \bB_{t+i} \bTe_t^T \rkv \|
\leq W_{\max} \max_{|i| \leq r} \|\bTe_{t+i} \bB_{t+i} \bTe_{t+i}^T - \bTe_t \bB_{t+i} \bTe_t^T  \|
\ees
where
$\|\bTe_{t+i} \bB_{t+i} \bTe_{t+i}^T - \bTe_t \bB_{t+i} \bTe_t^T  \| \leq 
[\|\bTe_{t+i} \| + \|\bTe_t \|] \, \|\bB_{t+i}\| \, \|\bTe_{t+i} - \bTe_t \|.
$
The last two inequalities together with \fr{eq:dif} imply \fr{eq:Del21t}.
Now, let us prove that 
\be \label{eq:Del22t}
\Del_{2,2,t} = \| \tilbP_{t,r}- \bP_t\| \leq 
\frac{L}{l!}\,  W_{\max}\, \al_n n\, \lkr \frac{r}{T} \rkr^\beta.
\ee
Let $j=1,2$ or 3 be determined by the value of $t$.
Denote 
$$
\bQ_{r,t} = |\calF_{r,j}|^{-1}\ \sum_{i\in \calF_{r,j}}  W^{(j)}_{r,l} (i) 
\lkr \bH_{t+i}- \bH_t \rkr.
$$
Observe that by Assumptions {\bf A1} and {\bf A4}, for any $k,k' = 1, \ldots, K$, using Taylor's expansion at 
$i=0$, one derives
\beqns
\bQ_{r,t} (k,k') & = &   |\calF_{r,j}|^{-1}\ \sum_{i\in \calF_{r,j}}  W^{(j)}_{r,l} (i) 
\lkv f\lkr \frac{t+i}{T};k,k' \rkr - f\lkr \frac{t}{T};k,k' \rkr \rkv  \\
 & = & \sum_{h=1}^l \, \frac{1}{h!}\,  f^{(h)} \lkr \frac{t}{T};k,k' \rkr \, 
\lkv \frac{1}{r}\ \sum_{i\in \calF_{r,j}} \lkr \frac{i}{T}\rkr^h W^{(j)}_{r,l} \rkv\\
& + & \frac{1}{|\calF_{r,j}|\, l!}\ \sum_{i\in \calF_{r,j}} W^{(j)}_{r,l}\, \lkr \frac{i}{T}\rkr^l\, 
\lkv f^{(l)} \lkr \frac{t}{T} + \xi;k,k' \rkr - f^{(l)} \lkr \frac{t}{T};k,k' \rkr \rkv,
\eeqns
where $|\xi| \leq r/T$. Due to Assumption  {\bf A1}, the first sum is equal to zero and 
\be  \label{eq:abs_delrt}
|\bQ_{r,t}  (k,k')| \leq \frac{1}{ |\calF_{r,j}|\, l!}\ \sum_{i\in \calF_{r,j}} \lkr \frac{i}{T}\rkr^l\,  
|W^{(j)}_{r,l} (i)| \, L |\xi|^{\beta - l} 
\leq \frac{L W_{\max}}{l!}\, \lkr \frac{r}{T}\rkr^\beta. 
\ee
Recall that $\bB_{t+i}- \bB_t = \al_n(\bH_{t+i}- \bH_t)$  and that the spectral norm  of a matrix is dominated by the $l_1$ norm.
Therefore, 
\beqns
\Del_{2,2,t}  & \leq & \al_n\, \max_{1 \leq j' \leq n} \ \sum_{j=1}^n |(\bTe_t  \bQ_{r,t} \bTe_t^T) (j,j')|\\
& \leq & \al_n\, \max_{k,k'}\, |\bQ_{r,t}  (k,k')| \ \max_{1 \leq j' \leq n} \ \sum_{k=1}^K \sum_{k'=1}^K \ 
\lkv \sum_{j \in G_{t, k}}  \bTe_t (j,k) \rkv\, \bTe_t (j',k')\\
& = & \al_n\, n \max_{k,k'}\, |\bQ_{r,t}  (k,k')|
\eeqns
Combination of the last inequality with \fr{eq:abs_delrt} yields \fr{eq:Del22t} while 
\fr{eq:Del2t}, \fr{eq:Del21t} and \fr{eq:Del22t} together complete the proof of the lemma. 
\\

\medskip


\noindent
{\bf Proof of Lemma \ref{lem:opt_error}. }
If $r^* =0$, then results of the Lemma follow directly from \cite{lei2015}. 
Consider $r \geq 1$. Then, 
\bes
\Del_t  (r) \leq C \lkv n \lkr \frac{r}{T} \rkr^\beta + 
\al_n  \, \sqrt{2 r n_{\max}\, s} + \sqrt{n \al_n /r} \rkv
\ees
where $C$ depends on $\tau, c_0, W_{\max},l,L$ and $\lam_{\max}$.
Denote 
\bes
F_1 (r) = n (r/T)^\beta,\quad F_2 (r) = \al_n  \, \sqrt{2 r n_{\max}\, s},
\quad F_3 (r) = \sqrt{n \al_n /r}.
\ees
It is easy to see that $F_1(r)$ and $F_2(r)$ are growing in $r$ while $F_3(r)$ is declining.
Therefore, the minimum is reached at the point $r$ where 
$F_1(r) + F_2 (r) \asymp \max(F_1(r),F_2 (r)) = F_3(r)$.
Observe that  $F_1(r) = F_3(r)$ if $r = r_1 = \lkr n^{-1} \al_n T^{2\beta}\rkr^{1/(2 \beta+1)}$
and  $F_2(r) = F_3(r)$ if $r = r_2 = \sqrt{ (\al_n n_{\max}\, s)^{-1}\, n}$. 
Moreover, $\max(F_1(r),F_2 (r))$ occurs at $r^* = \min(r_1, r_2)$ and we need $r^*$ to be an integer. 
Then, $\min_r (F_1 (r) + F_2(r) + F_3(r)) \asymp F_3(r^*)$ and plugging $r^*$ into $F_3(r)$,
we obtain \fr{opt_er}.
\\

\medskip


\noindent
{\bf Proof of Lemma \ref{lem:adapt}. }
Note that \fr{rhat} implies that for any $r_0 \geq \hr$ one has   
\be \label{lep1}
\| \hbP_{t,\hr}  - \hbP_{t, r_0}\| \leq 4\, C_{0,\tau}\, \sqrt{n\,\al_n/(r_0 \vee 1)}.
\ee
On the other hand, for any $r_0 > \hr$, there exists $\tilr < r_0$ such that 
\be \label{lep2}
\| \hbP_{t,\tilr}  - \hbP_{t,r_0}\| > 4\, C_{0,\tau}\, \sqrt{n\,\al_n/(\tilr  \vee 1)}.
\ee
Denote, for convenience, 
$\del_1 (r) = \|\bP_{t,r} - \bP_t\|$ and  
$\del_2 (r) =  4\, C_{0,\tau}\, \sqrt{n\,\al_n/(r  \vee 1)}$.
Since  $\del_1 (r)$ is growing in $r$ and $\del_2 (r)$ is decreasing in $r$,
there exists $r_0$ such that 
\bes  
\del_1(r_0) < \del_2(r_0), \quad \del_1(r_0+1) \geq \del_2(r_0+1)
\ees
Then, 
\beqns
\del_1(r^*) + \del_2(r^*)  & = & \min_r [\del_1(r) + \del_2(r)] \geq \min_r \max[\del_1(r), \del_2(r)]\\
& = & \max[\del_1(r_0+1), \del_2(r_0)] \geq \del_2(r_0) > [\del_1(r_0) + \del_2(r_0)]/2,
\eeqns
so that 
\be \label{del_rel}
\del_1(r_0) + \del_2(r_0) < 2[\del_1(r^*) + \del_2(r^*)] \leq 2[\del_1(r_0) + \del_2(r_0)].
\ee
Let $\Om_\tau$ be the set defined in Lemma \ref{lem:variance} and let $\om \in \Om_\tau$.
Now consider two cases: $\hr \geq r_0$ and $\hr < r_0$. 

If $\hr \geq r_0$, then  by \fr{lep1} one has
\beqns
\| \hbP_{t,\hr} - \bP_t\|   & \leq &   \| \hbP_{t,\hr}  - \hbP_{t, r_0}\|  + \| \hbP_{t,r_0} - \bP_t\| 
\leq 4\, C_{0,\tau}\, \sqrt{n\,\al_n/(r_0 \vee 1)} + \del_1 (r_0) + \del_2 (r_0)\\ 
& = & 5 \del_2 (r_0) +   \del_1 (r_0) \leq  5 [\del_1 (r_0) +   \del_2 (r_0)], 
\eeqns
so it follows from  \fr{del_rel} that
\be \label{eq:for1} 
\| \hbP_{t,\hr} - \bP_t\|   < 10\ \min_r [\del_1(r) + \del_2(r)].
\ee 
On the other hand, if $\hr < r_0$, then there exist $\tilr < r_0$ such that \fr{lep2} holds.
Therefore, due to $\del_1 (\tilr) < \del_1(r_0) < \del_2 (r_0) < \del_2(\tilr)$, obtain
\beqns
\|\hbP_{t,r_0} - \hbP_{t,\tilr}\| & \leq &   \|\hbP_{t,r_0} - \bP_t\| + \|\hbP_{t,\tilr} - \bP_t\|
\leq \del_1 (r_0) +   \del_2 (r_0) + \del_1 (\tilr) +   \del_2 (\tilr) < 4\,  \del_2 (\tilr)
\eeqns
 which contradicts \fr{lep2}. Hence, $\hr \geq r_0$ for $\om \in \Om_\tau$ and validity of 
Lemma \ref{lem:adapt} follows from \fr{eq:for1}.
\\

\medskip

\noindent
{\bf Proof of Theorem \ref{th:clus_er}. }
Recall that $\bP_t = \al_{n} \bTe_t \bH_t \bTe_t^T$ with  $\lam_{\min} (\bH_t) \geq C_{\lam}^{-1}$.
Since $\bTe_t^T \bTe_t = \bLam_t^2$, the diagonal matrix with $n_{t,1}, \cdots,  n_{t,K}$
on the diagonal,  $\bTe_t = \tilbU_t \bLam_t$ where $\tilbU_t = \bTe_t \bLam_t^{-1}$
is an orthogonal matrix. Therefore, in  \fr{tilRtexp} and \fr{Rtexp}
\bes
\lam_{\min}(\bP_t) \geq C_\lam^{-1}\,   \al_{n} \, n_{\min}.
\ees
Combining the last inequality, \fr{tilRtexp} and \fr{Rtexp} with Lemma \ref{lem:opt_error},
immediately obtain \fr{tilRtexp1} and \fr{Rtexp1}.
 \\

\medskip

\noindent
{\bf Proof of Proposition \ref{prop:clust_est}. }
Let $\Om_{t,\tau}$ be a set with $\Pr(\Om_{t,\tau}) \geq 1 - 4 n^{-\tau}$ where \fr{adap_err} hold.
Then, due to \fr{adap_err} and \fr{min_eig_con}, 
$\eps_t \leq 10 \,  \Del_t (r^*)/\lam_{K,t} \leq (4+w)^{-1}$.
Therefore, \fr{eig_ratios} and \fr{eig_exp_cond} yield  that
\bes
\frac{\hlam_{j+1,t}}{\hlam_{j,t}} \geq \frac{1 - (4+w)^{-1}}{1 + w + (4+w)^{-1}} > \frac{1}{3+w},\ j=1, \cdots, K-1,
\quad \frac{\hlam_{K+1,t}}{\hlam_{K,t}} \leq \frac{1}{3+w}
\ees
which completes the proof.
\\

\medskip


\subsection{Proofs of supplementary lemmas}
\label{sec:suppl_proofs}


\noindent
{\bf Proof of Lemma~\ref{lemma:lightpairs}. }
First, let us prove that for any $C>0$, one has 
\be \label{main_ineq_lem6}
\Pr\Big\{\sup_{\bx,\by\in\mathcal{T}}\Big|\sum_{(i,j)\in\mathcal{L}(\bx,\by)} x_i  y_j
[\hbP_{t,r} (i,j) - \bP_{t,r}(i,j)]  \Big| \geq C \sqrt{\frac{n\, C_{\al} \al_n}{r}}\Big\}
 \leq 2 \, e^{-n\, \lkr \frac{C^2}{4\,W_{\max}^2 + 4C\, W_{\max}/3}- \log 14\rkr}
\ee
Denote $u_{ij}=  x_i y_j \, \II(|x_i y_j|\leq  \sqrt{C_{\al} \al_n r/n})+ x_j y_i I(|x_j y_i|\leq  \sqrt{C_{\al} \al_n r/n})$. 
Consider 
$$
S = \sum_{(i,j)\in\mathcal{L}(\bx,\by)} x_i y_j\,(\hbP_{t,r}(i,j) - \bP_{t,r}(i,j))  = 
\frac{1}{|\calF_r|}\  \sum_{1\leq i<j\leq n} \ \sum_{k \in \calF_r} u_{ij}\, W_{r,l}(k) (\bA_{t+k}(i,j)-\bP_{t+k}(i,j)).
$$
Note that the right-hand side  is the sum  of $n(n-1) |\calF_r| $ independent variables 
$$
\xi_{i,j,k} = u_{ij} W_{r,l}(k)(\bA_{t+k}(i,j)-\bP_{t+k}(i,j))/|\calF_r|
$$  
with zero means and absolute values bounded by $|\xi_{i,j,k}| \leq 2 W_{\max} \sqrt{C_{\al} \al_n/n\, |\calF_r|}$, due to  
$|u_{ij}| \leq 2\sqrt{C_{\al} \al_n r/n}$ and $|\bA_{t+k}(i,j)-\bP_{t+k}(i,j)| \leq 1$. 
Applying Bernstein's inequality and using , the fact that $\sum_{i<j}u_{ij}^2\leq 2$  
(as proved in the end of Section 3 in the Supplementary material of \cite{lei2015}), obtain
\begin{align*}
  \Pr\left\{\sup_{\bx,\by\in\mathcal{T}}\left|   \sum_{1\leq i<j\leq n} \ \sum_{k \in \calF_r} \xi_{i,j,k} \right| 
\geq C\, \sqrt{\frac{n\, C_{\al} \al_n}{r}} \right\} 
& \leq    2\,\exp \left(-\frac{ \frac{C^2 n C_{\al} \alpha_n}{2r} } 
{ \frac{2\, C_{\al}  W_{\max}^2 \al_n}{|\calF_r|} +
  \frac{2 W_{\max}}{3}\, \sqrt{\frac{C_{\al} \al_n}{n  |\calF_r|}}\ C\, \sqrt{\frac{n C_{\al} \alpha_n}{r}} } \right)\\
%
& \leq    2\, \exp\left(-\frac{C^2n}{4\,  W_{\max}^2+\frac{4C}{3}W_{\max}}\right),
\end{align*}
since $|\calF_r|\geq 1 + r$. Using the fact that cardinality  $|\mathcal{T}| \leq \exp(n \log 14)$
(see Section 3 in the Supplementary material of \cite{lei2015} with $\delta=1/2$), obtain \fr{main_ineq_lem6}.
In order to complete the proof, observe that inequality \fr{Ctau1} guarantees that the right-hand side in 
\fr{main_ineq_lem6} is bounded by $2 n^{-\tau}$. 
 \\

\medskip


\noindent
{\bf Proof of Lemma~\ref{lemma:heavypairs1}. }
First, we shall prove that for any $c_1>1$, one has
\be \label{ineq_main_lem7}
\Pr \lfi \max_{1 \leq i \leq n} d_{t,r}(i) \leq c_1\, C_{\al}\, n\,  \al_n\, r \rfi \geq
 1-  n^{1-\frac{3(c_1-1)^2c_0\, r}{6\,W_{\max}^2 + 2 W_{\max}(c_1-1)}}.
\ee
For a fixed node $i$, using Bernstein's inequality and $C_{\al} \al _n n \geq c_0 \log n$, obtain
\begin{align*}
&\Pr(d_{t,r}(i)> c_1 n C_{\al} \alpha_n |\calF_r|)\leq \Pr\Big(\sum_{j=1}^n\sum_{k\in \calF_{r}} W_{r,l}(k) 
[\bA_{t+k}(i,j)-\bP_{t+k}(i,j)]\leq (c_1-1) C_{\al} n \alpha_n |\calF_r| \Big)\\
\leq& \exp\Big(-\frac{\frac{1}{2}(c_1-1)^2 C_{\al}^2\,  n^2 \alpha_n^2 |\calF_r|^2}
{C_{\al} |\calF_{r}| W_{\max}^2   n   \alpha_n+ \frac{1}{3} W_{\max}(c_1-1) C_{\al} n \alpha_n |\calF_r|}\Big)
\leq n^{-\frac{3\,(c_1-1)^2c_0 |\calF_r|}{6   W_{\max}^2+2 W_{\max}(c_1-1)}}.
\end{align*}
Taking the union bound over $i=1, \cdots, n$, derive \eqref{ineq_main_lem7}.

In order to complete the proof, note that inequality \eqref{Ctau2} guarantees that the 
right hand side of \eqref{ineq_main_lem7} is bounded below by $1 - n^{-\tau}$ due to
$\max(r, 2) \leq |\calF_r| \leq 3 r$ for $r \geq 1$. 
\\

\medskip


\noindent
{\bf Proof of Lemma~\ref{lemma:heavypairs2}. }
First, if we divide the weights $W_{r,l}(k)$ into two groups: 
$\mathcal{K}_1=\{k\in \calF_{r}: W_{r,l}(k)> 0\}$ and  $\mathcal{K}_2=\{k\in \calF_{r}: W_{r,l}(k)\leq 0\}$.
Define
\[
Y_{ijk}=I(\bA_k(i,j)=1)\cdot \II(k\in \mathcal{K}_1).
\]
Then each $Y_{ijk}$ is a Bernoulli random variable with expectation  $\bP_k(i,j)\cdot I(k\in \mathcal{K}_1)$, and by definition
\begin{equation}\label{eq:lemma8_1}
e_{t,r}(I,J)=\sum_{k\in \calF_{r}}W_{r,l}(k) e_{t+k}(I,J)\leq\sum_{k\in\mathcal{K}_1}W_{r,l}(k) e_{t+k}(I,J)=
\sum_{k\in \calF_{r}} \sum_{i\in I} \sum_{j\in J}\, W_{r,l}(k)Y_{ijk}.
\end{equation}
Applying Lemma~\ref{Lemma:sum_poisson} below with $X_i$ replaced by $Y_{ijk}-\mathbb{E} Y_{ijk}$, 
 $w_{\max}$ replaced by $W_{\max}$,   $p_{\max}$ replaced by $\alpha_n$, $k$ replaced by $t$
 and $n=|\calF_{r}||I||J|$, obtain for $t\geq \max(e^{3W_{\max}},2)$: 
\begin{equation}\label{eq:lemma8_2}
\Pr\lfi\sum_{k\in \calF_{r}, i\in I, j\in J}W_{r,l}(k)[Y_{ijk}-\mathbb{E} Y_{ijk}] > t\ \bar{\mu}(I,J) \rfi
\leq \exp\lkv -\frac{(t+1)\ln (t+1) C_{\al} \alpha_n|\calF_{r}||I||J|} {2W_{\max}}\rkv.
\end{equation}
Since $\mathbb{E} Y_{ijk}<C_{\al} \alpha_n$, application of  \eqref{eq:lemma8_1}  and \eqref{eq:lemma8_2}  
with $t\geq \max(e^{3W_{\max}},2)+W_{\max}$ yields
\begin{align*}
& \Pr\left\{e_{t,r}(I,J)> t\ \bar{\mu}(I,J)\right\}
\leq \Pr\left\{\sum_{k\in \calF_{r}} \sum_{i\in I} \sum_{j\in J}\, W_{r,l}(k)Y_{ijk} > t\ \bar{\mu}(I,J) \right\} \\
\leq & \Pr\left\{\sum_{k\in \calF_{r}} \sum_{i\in I} \sum_{j\in J}\, 
W_{r,l}(k)[Y_{ijk}-\mathbb{E} Y_{ijk}] > (t-W_{\max}) \bar{\mu}(I,J) \right\}\\
\leq& \exp\Big[-\frac{(t+1-W_{\max})\ln (t+1-W_{\max}) C_{\al} \alpha_n|\calF_{r}||I||J|}
{2W_{\max}}\Big]= \exp\Big[-\frac{(t+1-W_{\max})\ln (t+1-W_{\max})  \bar{\mu}(I,J)} {2W_{\max}}\Big].
\end{align*}
Using the fact that   $s>\max(a,2)$ implies $(s+a)\ln(s+a)\leq (s+a)[\ln s+\ln\max(a,2)]\leq 4s \ln s$,
and setting $a = W_{\max}-1$ and $s = t+1-W_{\max}$, obtain that  
$$
\frac{(t+1-W_{\max})\ln (t+1-W_{\max})  \bar{\mu}(I,J)} {2W_{\max}} \geq 
\frac{t\ln t  \bar{\mu}(I,J)}{8W_{\max}}
$$
provided $t+1-W_{\max} > \max(W_{\max}-1,2)$. In summary, if $t > C_{\tau,5}$ where 
$C_{\tau,5} = \max(e^{3W_{\max}},W_{\max}-1,2)+W_{\max}$, then 
\[
\Pr\left\{e_{t,r}(I,J)> t \bar{\mu}(I,J)\right\}\leq  \exp\Big[-\frac{t\  \ln t\   \bar{\mu}(I,J)}
{8W_{\max}}\Big].
\]
The rest of the proof repeats the proof of Lemma 4.2 in \cite{lei2015} 
(start from the fourth paragraph in Section 4.2, note that the constant $8$ is replaced by $C_{\tau,5}$, $\frac{1}{2}$ 
in the exponent is replaced by $\frac{1}{8W_{\max}}$, $c$, $c_1$, $c_2$ and $c_3$ are replaced, respectively,  by 
$\tau$, $3(W_{\max}C_{\tau,2} +1)$, $C_{\tau,3}$ and $C_{\tau,4}$). 
In particular, by following that calculation, we can show that if
$C_{\tau,3} =\max\{3(W_{\max}C_{\tau,2} +1), C_{\tau,5}\}$  and $C_{\tau,4}$ is  chosen so that 
$C_{\tau,4}/(8W_{\max})- 6= \tau$, that is  $C_{\tau,4}=8W_{\max}(\tau+6)$,
then Lemma~\ref{lemma:heavypairs2} is valid. To complete the proof, note that $\max(W_{\max}-1, 2) \leq 3(W_{\max}C_{\tau,2} +1)$.
\\

\medskip


\begin{lemma}\label{Lemma:sum_poisson}
Let $\{X_i\}_{i=1}^n$ be random variables such that $\Pr(X_i=1-p_i)=p_i$, 
$\Pr(X_i=-p_i)=1-p_i$ for some $p_i>0$. Let $X=\sum_{i=1}^n w_iX_i$, 
$p=\frac{1}{n}\sum_{i=1}^np_i$, $p_{\max}=\max_{1\leq i\leq n} p_i$, 
$w_{\max}=\max_{1\leq i\leq n} w_i$, then for $k>\max(e^{3w_{\max}},2)$, \[
\Pr(X\geq k p_{\max}n)<   e^{-(k+1) p_{\max}n\ln (k+1)/2w_{\max}}.\]
\end{lemma}

\medskip

\noindent
{\bf Proof. } 
$E(e^{\lambda X_i})=p_ie^{w_i (1-p_i)\lambda}+(1-p_i)e^{-w_i p_i \lambda}$.
Following the same proof as Lemma 2.1.8 in (Alon and Spencer, 2004), we have
\[
E(e^{\lambda X})\leq e^{-\sum_{i=1}^nw_ip_i \lambda}[pe^{w_{\max}\lambda}+(1-p)]^n.
\]
Let $\lambda=\ln[1+a/pn]/w_{\max}$, then using $(1+a/n)^n\leq e^a$, its right hand side is bounded by
\[
  e^{ a-\sum_{i=1}^nw_ip_i \ln[1+a/pn]/w_{\max}}.
\]
As a result, 
\[
\Pr(X\geq a)<e^{-a\lambda}E(e^{\lambda X})\leq e^{ a-(\sum_{i=1}^nw_ip_i+a) \ln[1+a/pn]/w_{\max}}. 
\]
Let $a=k p_{\max}n$, then
\begin{align*}
&\Pr(X\geq k p_{\max}n)<   e^{ k p_{\max}n-(\sum_{i=1}^nw_ip_i+k p_{\max}n) \ln[1+k]/w_{\max}}
<e^{k p_{\max}n(1-\ln (k+1)/w_{\max})} \\<&e^{-(k+1) p_{\max}n\ln (k+1)/2w_{\max}},
\end{align*}
where the last inequality holds when $k>\max(e^{3w_{\max}},2)$.


\bibliographystyle{abbrv}
\bibliography{citations}

\end{document}